\documentclass[aps,prl,preprint,superscriptaddress]{revtex4-1} 

\usepackage{amsmath,amssymb,amsfonts}
\usepackage{graphicx}
\usepackage{textcomp}

\usepackage{pstricks}
\usepackage{pst-3d}
\usepackage{pst-eps}

\definecolor{mypink1}{rgb}{0.85, 0.85, 0.85}
\definecolor{mypink2}{rgb}{0.75, 0.75, 0.75}
\definecolor{mypink3}{rgb}{0.5, 0.5, 0.5}
\definecolor{mypink4}{rgb}{0.35, 0.35, 0.35}

\begin{document}

\title{Exceptional points in anisotropic planar microcavities}

\author{Steffen Richter}
\email{steffen.richter@physik.uni-leipzig.de}
\affiliation{Universit\"at Leipzig, Institut f\"ur Experimentelle Physik II, Linn\'estr. 5, 04103 Leipzig, Germany}

\author{Tom Michalsky}
\affiliation{Universit\"at Leipzig, Institut f\"ur Experimentelle Physik II, Linn\'estr. 5, 04103 Leipzig, Germany}

\author{Chris Sturm}
\affiliation{Universit\"at Leipzig, Institut f\"ur Experimentelle Physik II, Linn\'estr. 5, 04103 Leipzig, Germany}

\author{Bernd Rosenow}
\affiliation{Universit\"at Leipzig, Institut f\"ur Theoretische Physik, Br\"uderstr. 16, 04103 Leipzig, Germany}

\author{Marius Grundmann}
\affiliation{Universit\"at Leipzig, Institut f\"ur Experimentelle Physik II, Linn\'estr. 5, 04103 Leipzig, Germany}

\author{R\"udiger Schmidt-Grund}
\affiliation{Universit\"at Leipzig, Institut f\"ur Experimentelle Physik II, Linn\'estr. 5, 04103 Leipzig, Germany}


\date{Sep 2016}

\begin{abstract}

Planar microcavities allow the control and manipulation of spin-polarization, manifested in phenomena like the optical spin Hall effect due to the  intrinsic polarization mode splitting. 
Here, we study a transparent microcavity with broken rotational symmetry, realized by aligning the optical axis of a uniaxial cavity material in the cavity plane. 
We demonstrate that the in-plane optical anisotropy gives rise to exceptional points in the dispersion relation, which occur pair-wise, are circularly polarized, and are cores of polarization vortices. 
These exceptional points are a result of the non-Hermitian character of the system, and are in close relationship to singular optical axes in absorptive biaxial systems. 

\end{abstract}

\maketitle

Optical microcavities are widely used structures to tailor light-matter interactions \cite{deng2010}, with much emphasis on the control and manipulation of spin-polarization of cavity photons and polaritons \cite{martin02,amo2010,dreismann2016,bhattacharya16}. 
The intrinsic  splitting of modes with different linear polarizations in isotropic planar microcavities can be described by a pseudo-magnetic field, and gives rise to phenomena like the optical spin Hall effect \cite{oshe1,maragkou,oshe2,Kammann2012}. 
An additional external magnetic field can help to tune the polarization characteristics \cite{larionov10,Morina2013}. 
Some of the polarization effects are related to spontaneous symmetry breaking \cite{levrat10,ohadi2012,schmitt2016}, with only indirect control over them. 
In dynamical settings, the polarization displays  the emergence of vortices in real or momentum space \cite{liew08,dufferwiel2015,cilibrizzi16}. 
This type of spin-momentum coupling is described by singular optics, which studies the effect of vortex centers on polarization and topology of an optical system \cite{nye1983,dennis2009,soskin2001}. 

Spin vortices can also occur in the form of circular polarized points ($C$-points). Under conditions described below, $C$-points can be realized in the form of exceptional points (EPs), at which not only energies (eigenvalues) but also eigenstates of a system are degenerate \cite{heiss2000, heiss2012}. EPs  occur  in a variety of  systems \cite{dembowski2001, lee09, wiersig2011, dietz2011, kim2013, hahn16, peng2016} described by non-Hermitian Hamiltonians. In optically biaxial crystals, degeneracies with a two-dimensional polarization eigenspace occur along optical axes and are called diabolical points. 
However, in absorptive media, both optical axes split into two singular axes \cite{voigt1902,berry2003,sturm2016}, thus realizing EPs \cite{cao2015, berry2004, heiss2004} with coalescing eigenstates. Here, each eigenstate represents one circular polarization \cite{heiss2001,dennis2002,freund2002}, such that only left or right circularly polarized light is allowed to propagate in a given direction ("Voigt wave"). In this way, optically biaxial, dissipative structures provide a path towards spin-momentum coupling and spatial separation of circular polarizations. 

In this letter, we discuss a transparent planar microcavity with broken rotational symmetry, realized by aligning the optical axis of a uniaxial cavity material in the cavity plane. 
The  photonic modes of such a cavity realize a dissipative system with orthorhombic symmetry. Hence, the conditions  for EPs are satisfied: optical biaxiality and non-Hermiticity. The experimental feasibility of such structures has already been demonstrated \cite{tao2015,rossbach2011,jesus}. Specifically, we propose a ZnO-based microcavity with dielectric Bragg reflectors (DBRs) consisting of  layered pairs of Al$_2$O$_3$ and Y-stabilized ZrO$_2$. The advantages of our proposal are i) the use of non-absorbing materials, i.e.~all injected energy is emitted in terms of photons, ii) a microcavity which can be incorporated into opto-electronic devices, and iii) highly tunable characteristics of the microcavity due to the free choice of materials and geometries. 

Photons in a planar microcavity have free wavevector components $\vec{k}_{||}=(k_x,k_y)^\text{T}$ in the cavity plane, and a quantized component $k_z$ perpendicular to the cavity plane. 
Using a $4\times4$ transfer matrix approach \cite{berreman,schubertprb,steffen}, we compute the photonic cavity modes described by the electromagnetic field amplitudes along the $x$ and $y$ direction $(\mathcal{E}_{x},\mathcal{E}_{y})^\text{T}$ as well as their complex mode energies $\tilde{E}=E-i\gamma$, depending on the (real) in-plane wavevector $\vec{k}_{||}$. Here, $\gamma$ is the half width at half maximum of the mode, representing photonic losses. Modes are found as matrix roots of the generalized mode condition: 
\begin{equation}
\hat{J}_{trans}^{-1}(\tilde{E},\vec{k}_{||})\begin{pmatrix} \mathcal{E}_{x} \\ \mathcal{E}_{y} \end{pmatrix}_\text{out} = \begin{pmatrix} 0 \\ 0 \end{pmatrix}_\text{in}
\label{eq:condition}
\end{equation} 
where $\hat{J}_{trans}$ is the transmission Jones matrix for the microcavity. We restrict the values of $|\vec{k}_{||}|$ to be within the vacuum light cone, i.e. $|\vec{k}_{||}|<n_aE/\hbar c_0$ with ambient refractive index $n_a=1$. $\hbar$ and $c_0$ are the reduced Planck constant and vacuum speed of light, respectively. 
If the cavity has a mirror symmetry with regards to the $x$-$y$-plane, then $\hat{J}_{trans}$ and its inverse are symmetric matrices. 
Denoting the diagonal elements by ${\rm diag}(\hat{J}_{trans}^{-1})=(a,c)$ and the off-diagonal element by $b$ ($a,b,c\in\mathbb{C}$), the eigenvalues of $\hat{J}_{trans}^{-1}$ are given by $\lambda_\pm = (a+c \pm \sqrt{(a-c)^2  + 4 b^2})/2$. In general, eigenmodes of the cavity are characterized by one eigenvalue $\lambda =0$. At an exceptional point two such eigenvalues merge, such that both eigenvalues $\lambda_\pm=0$. 
As a consequence, one finds the conditions $(a-c)^2 = - 4 b^2$ and  $c=-a$, which together imply $b = \pm i a$. Then, there is only one eigenvector $\vec{v} = (\mathcal{E}_{x}^\text{EP},\mathcal{E}_{y}^\text{EP})^\text{T} = \mathcal{E}_{0}^\text{EP}(1,\pm i)^\text{T}$, which corresponds to circularly polarized light according to Eq.\,\ref{eq:pscomponentdefinition} below. We thus have established that exceptional points are indeed $C$-points. 

As exemplary model system we investigate a microcavity consisting of an optically uniaxial $\lambda/2$ cavity layer surrounded symmetrically by DBRs. We neglect the effect of a substrate, keeping the structure mirror-symmetric with respect to the cavity plane. 
The structure parameters we use are idealized parameters of a ZnO-based microcavity with DBRs consisting of 15 layer pairs Al$_2$O$_3$ and Y-stabilized ZrO$_2$ each \cite{helena}. 
For the cavity dielectric function a positive birefringence of 2.7\% is assumed ($n_\perp=2.20, n_{||}=2.26$, neglecting dispersion) which is in the order of typical values for ZnO or GaN. 
We consider different detunings between twice the cavity optical thickness $n_cd_c$ ($n_c=n_{\perp,||}$) and the DBR central wavelength which we set to $\lambda_c=4n_{j\text{DBR}}d_{j\text{DBR}}=496$\,nm (2.5\,eV). The refractive indices of the DBR materials are chosen to be constant as 1.8 and 2.2, respectively, which yields a Bragg stopband width of about 0.5\,eV.

In isotropic microcavities, each cavity photon mode is split into a transverse electric ($TE$) and a transverse magnetic ($TM$) polarized one \cite{mcbook}. In anisotropic cavities, in general, polarization conversion occurs. We find that, irrespective of the orientation of the optical axis, the $\lambda/2$-cavity photon is generally split into two modes. Technical details of the computational approach are explained in the supplemental material \cite{suppl}. 
This is in contrast to calculations which consider modes in the basis of $TE$ and $TM$ polarization separately and, hence, lack of full polarization treatment \cite{sturm2011}. 
As soon as the rotational symmetry of the microcavity is broken, i.e. the optical axis is not oriented parallel to the surface normal, these modes become mostly elliptically polarized. 
No qualitative difference is found if the optical axis is oriented inside the cavity plane or tilted against it. Thus, we restrict the detailed discussion to the first case. We choose the optical axis to be always aligned along the $y$-axis of the laboratory coordinate system. We further define modes 1 and 2 such that it holds $E_2>E_1$. 

\begin{figure}
	\centering
	\includegraphics[width=.8\textwidth]{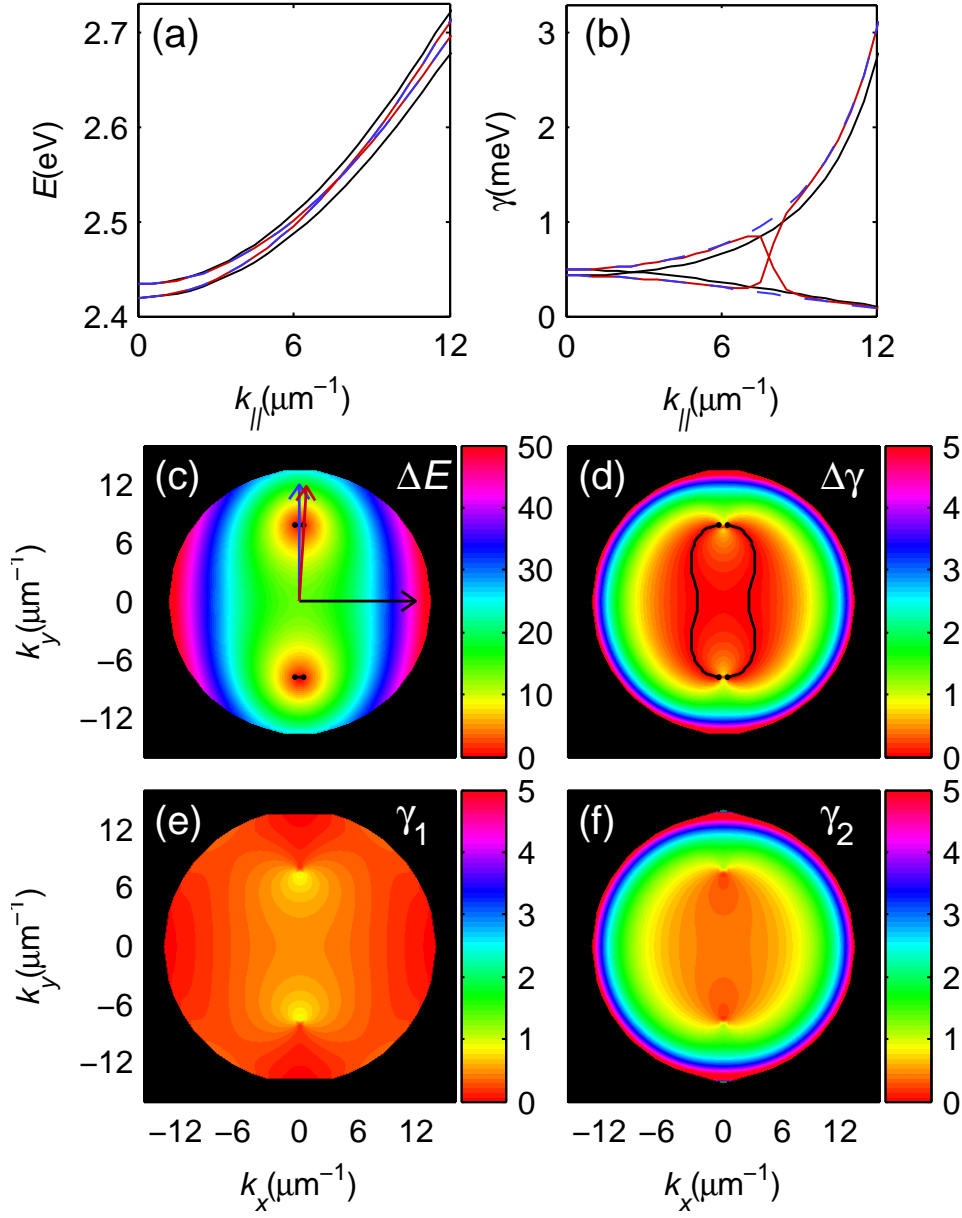}
	\caption{Cavity photon modes for $d_c=130$nm, i.e. $n_cd_c>\lambda_\text{DBR}/2$, and the cavity layer's optical axis along $y$: (a) mode energy $E$ and (b) broadening $\gamma=HWHM$ for three different propagation directions which are depicted in (c): along $x$ (black), $y$ (blue dashed) and crossing an EP (red). Differences of the (c) mode energies and (d) mode broadenings in the momentum space. The black contours illustrate degeneracy of $E$ and $\gamma$, respectively. Both are simultaneously degenerate at the EPs (black dots). Mode broadening $\gamma$ of (e) mode 1 and (f) mode 2 showing the discontinuities in the momentum space. For all colorbars, the unit is meV. The shape of the edge of the light cone differs for the two modes due to the optical anisotropy.}
	\label{fig:uebersicht130nm}
\end{figure}

Figure \ref{fig:uebersicht130nm} gives an overview of the cavity photon modes of the uniaxial microcavity for the case that the optical thickness of the cavity is larger than half the central DBR wavelength. 
While the real part $E$ of the complex mode energies (Fig.\,\ref{fig:uebersicht130nm}\,(a)) of both modes degenerates at certain $\vec{k}_{||}$-values for propagation along the optical axis, the imaginary part $\gamma$ does not (Fig.\,\ref{fig:uebersicht130nm}\,(b)). However, there are four finite $\vec{k}_{||}$-values where the complex mode energies $\tilde{E}$ degenerate, establishing EPs. Those occur pair-wise for propagation directions slightly off from the optical axis. 
As can be seen in Fig.\,\ref{fig:uebersicht130nm}\,(e) and (f), discontinuities occur for the mode broadening in the momentum space along the trajectory of $E_1=E_2$, also visible in the red line of Fig.\,\ref{fig:uebersicht130nm}\,(b). Here, the broadening values of both modes merge continuously into each other. Consequently, changing the mode assignment can resolve this discontinuity locally. 
Such a mode exchange and mutual crossing/anti-crossing of real and imaginary part of the mode energy is characteristic for encircling single EPs in the complex energy or momentum space \cite{heiss2000,dembowski2001,wiersig2011}. 

\begin{figure}
	\centering
	\includegraphics[width=.8\textwidth]{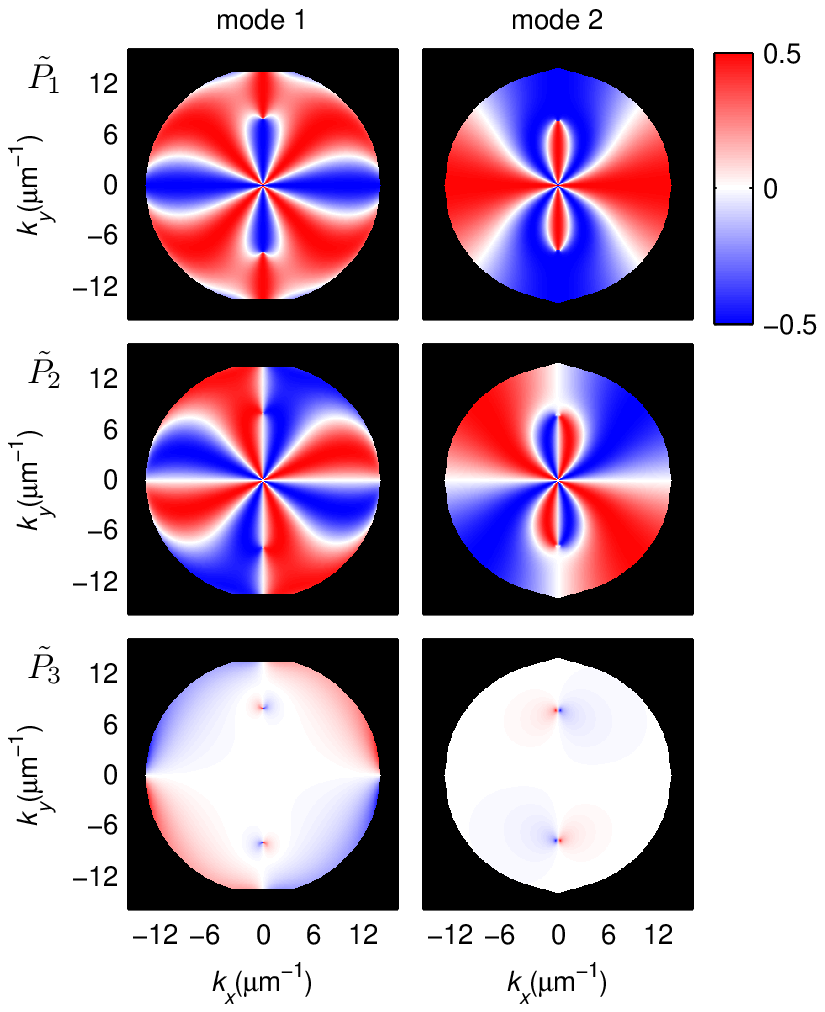}
	\newpage
	\caption{Mode polarization in terms of the pseudospin according to Eq.\,\ref{eq:pscomponentdefinition} for the same microcavity as in Fig.\,\ref{fig:uebersicht130nm} (optical axis along $y$). The EPs appear as circularly polarized spots and polarization discontinuities occur for $\tilde{P}_1$ at those $\vec{k}_{||}$ where the real part of the complex mode energy $E$ is degenerate (cf. Fig.\ref{fig:uebersicht130nm}).} 
	\label{fig:polarisationsuebersicht130nm}
\end{figure}

The mode polarization can be described using the polarization pseudospin vector $\vec{\tilde{P}}$ which corresponds to the state's Bloch vector with circularly polarized basis states. 
Its SO(3) components are given by
\begin{equation}
\begin{pmatrix} \tilde{P}_1 \\ \tilde{P}_2 \\ \tilde{P}_3 \end{pmatrix} = 
\frac{1}{2} \begin{pmatrix} |\mathcal{E}_{x}|^2-|\mathcal{E}_{y}|^2 \\ \mathcal{E}_{x}\mathcal{E}_{y}^*+\mathcal{E}_{x}^*\mathcal{E}_{y} \\ i(\mathcal{E}_{x}\mathcal{E}_{y}^*-\mathcal{E}_{x}^*\mathcal{E}_{y}) \end{pmatrix} 
\label{eq:pscomponentdefinition}
\end{equation}
with $|\mathcal{E}_{x}|^2+|\mathcal{E}_{y}|^2=1$. 
While $\tilde{P}_1$ represents the difference of the linear polarization contributions along $x$ and $y$, $\tilde{P}_2$ expresses the same for the linear polarizations tilted $+45^\circ$ and $-45^\circ$, and $\tilde{P}_3$ indicates the difference between right- and left-circular contributions. This is similar to the polarization Stokes vector but for the pseudospin it holds $|\vec{\tilde{P}}|=1/2$. 
Figure \ref{fig:polarisationsuebersicht130nm} illustrates the pseudospin components in the $\vec{k}_{||}$-space. The generally elliptic polarization is apparent and the mode degeneracies can be clearly identified as circularly polarized spots, which are referred to as "$C$-points" \cite{bliokh2008,cardano2013,fadeyeva2010,flossmann2006}. At these points, the respective signs of $\tilde{P}_3$ are the same for both modes, emphasizing that not only the eigenvalues but also the eigenmodes are degenerate at the EPs, in contrast to diabolic points. 
For each of the two pairs of EPs in the $\vec{k}_{||}$-space, it holds that one point carries right, the other one left circular polarization, yielding a trivial net topology. 
$C$-points are are surrounded by lines of linear polarization ("$L$-lines"). Here, the $k_x$ and $k_y$ axes serve as $L$-lines. Apart from them, the modes are generally not orthogonal to each other, i.e. $\mathcal{E}_{x,1}\mathcal{E}_{x,2}^*+\mathcal{E}_{y,1}\mathcal{E}_{y,2}^*\neq0$. This is in fundamental contrast to rotational symmetric (isotropic) microcavities and hinders the application of Zeeman-like pseudospin-Hamiltonians \cite{amo09,oshe1}. 
The symmetry of $\vec{\tilde{P}}$ (as well as the one of $\tilde{E}$) in the $\vec{k}_{||}$-space reflects the orthorhombic symmetry of the structure. While the rotational symmetry is broken, there is still in-plane inversion symmetry. 
$z$ inversion, i.e. considering forwards vs. backwards traveling modes, changes the signs of $\tilde{P}_2$ and $\tilde{P}_3$ as a consequence of the definition of a right-handed polarization coordinate system with respect to the propagation direction. 

\begin{figure}
	\centering
	\includegraphics[width=.8\textwidth]{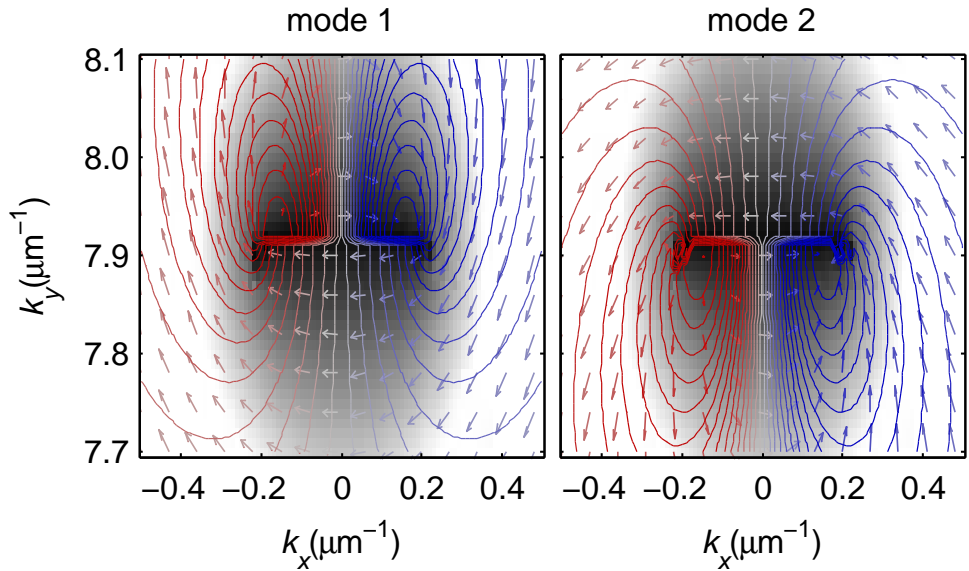}
	\caption{Detailed view of the polarization of the two modes in the momentum space region around a pair of EPs for the same situation as in Fig.\,\ref{fig:polarisationsuebersicht130nm}. The arrows represent the linear polarization vector, i.e. $\tilde{P}_1$ pointing along the $k_{x}$ axis and $\tilde{P}_2$ pointing along the $k_{y}$ axis. The circular polarization component $\tilde{P}_3$ is represented by the arrows' colors and also by the contour lines. The color scale is the same as in Fig.\,\ref{fig:polarisationsuebersicht130nm}. The gray-scaled background represents the mode splitting $E_2-E_1$ which is zero (black) along the connection line between the two EPs.}
	\label{fig:polarisationspunktdetail}
\end{figure}

Similarly to the imaginary part of the mode energies, a discontinuity in the momentum space occurs for $\tilde{P}_1$ along the line of degenerate real part of the mode energies. Again, this discontinuity can be overcome by exchanging the modes upon crossing this line. 
A close-up of the polarization near the EPs (Fig.\,\ref{fig:polarisationspunktdetail}) shows that the linear polarization component $(\tilde{P}_1,\,\tilde{P}_2)^\text{T}$ performs a vortex of winding number $+1$ for a closed path around a pair of EPs in the momentum space similar to what \textcite{voigt1902} described in 1902 for absorptive biaxial crystals near the singular optical axes. 
Each of the EPs corresponds to a vortex kernel of winding number $+1/2$, which is their topological charge \cite{angelsky2002}. The sign of the vortices is the same for the two modes and it is also the same for the EPs at opposite $\vec{k}_{||}$. 
The polarization-momentum coupling at the EPs can be understood similar to the optical Hall or spin Hall effect \cite{onoda04,bliokh08,oshe1}: For light propagating here along the optical axis, lateral scattering yields spatial separation of right and left circular polarizations. This mechanism is robust against scattering between mode 1 and 2. 

The EPs described above are signatures of singular optical axes of the effectively orthorhombic re\-so\-na\-tor structure. As in respective homogeneous media, four such axes occur \cite{sturm2016}. While we observe separate lines of degenerate real and imaginary part of the mode energies in momentum space, such lines can be found for the real and imaginary part of the complex refractive index in absorptive biaxial media. 
The general occurrence of effective singular optical axes in composite media of two uniaxial, absorbing materials has already been shown using effective medium theory \cite{mackay2000}. However, the description of multi-layer systems like the planar microcavity considered here, is beyond this approach. 

Furthermore, the setup studied here is related to a non-Hermitian matrix operator \cite{suppl}. The non-Hermiticity reflects the fact that the system is dissipative, which is also the key for the appearance of EPs in the first place \cite{berry2004,heiss2001,heiss2004}. 
For singular optical axes in biaxial media, the origin of the non-zero imaginary part of the complex mode energies or wavevectors is absorption, which limits applications. The microcavity, however, is non-absorbing but features non-zero mode broadening, and thus allows photon escape and hence practical utilization. 

\begin{figure}
	\centering
	\includegraphics[width=.8\textwidth]{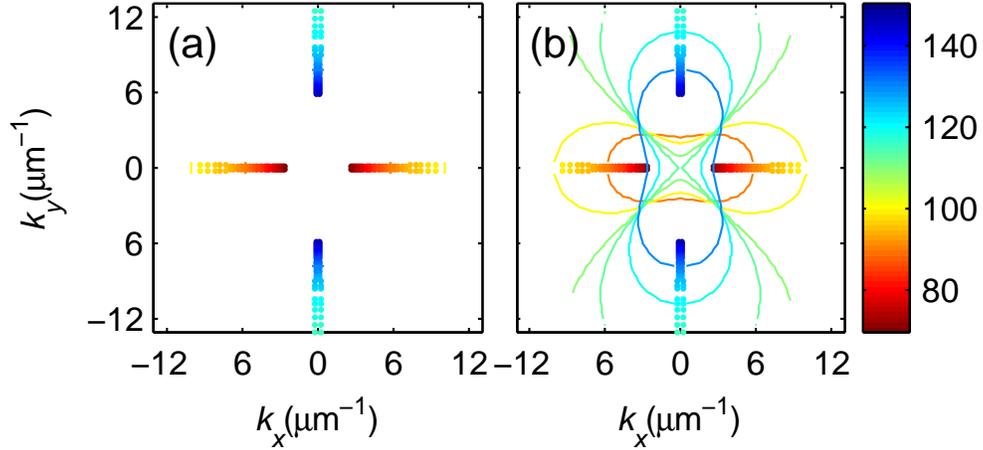}
	\caption{Lines: Paths in the momentum space of degenerate (a) real part of the complex mode energy $E$ and (b) imaginary part $\gamma$ for selected cavity thicknesses $d_c$, given by the color bar in nm. The EPs are shown in both plots as dots. They occur only within the vacuum light cone if the cavity is detuned with respect to the DBRs. Note that the interruptions of the "$\gamma$-degeneracy lines" are exactly the lines where $E$ is degenerate.}
	\label{fig:entartungspunkttracking}
\end{figure}

A further requirement for the manifestation of EPs is a detuning between cavity and (half the) central DBR wavelength: 
If the optical cavity thickness $n_cd_c$ is sufficiently larger than half the DBR central wavelength, the EPs occur for propagation nearly along the optical axis of the cavity layer; if $n_cd_c<\lambda_\text{DBR}/2$, they appear for propagation nearly perpendicular to it. (For negative birefringence we find the opposite behaviour.) Otherwise they are pushed out of the light cone. A condition for the occurrence of mode degeneracies can be deduced from \textcite{sturm2011} for the cases $c||x$ and $c||y$, where the separation of $TE$ and $TM$ polarization is correct. 
Figure \ref{fig:entartungspunkttracking} illustrates the dependence of the degeneracies on the thickness of the cavity layer. Further computed polarization patterns can be found in the supplemental material \cite{suppl}. 
While singular optical axes in naturally biaxial bulk crystals cannot be designed, angular and energetic position of EPs of the microcavity can be engineered via geometry and material choice. 

In summary, we have performed a numerically exact modelling  of plane wave modes in an optically anisotropic planar $\lambda/2$ microcavity, and have found  two generically elliptically polarized, non-orthogonal cavity photon states. 
For special in-plane momenta, these states  become degenerate in both energy and broadening, realizing 
 two pairs of exceptional points in momentum space. These points are circularly polarized, and serve as vortex centers for the linear polarization vector. The occurrence of exceptional points can be controlled by varying the cavity thickness relative to the central wavelength of the surrounding Bragg mirrors. 
Hence, a purely photonic but effectively biaxial planar microcavity is sufficient to observe polarization vortices. 
In a next step, the system analyzed here  can be extended, for instance  by coupling the cavity to an electronic system with e.g. excitonic resonances yielding anisotropic polaritonics. 

We thank Lennart Eicke Fricke for advices on transfer and scattering matrices and Christian Kranert for helpful technical hints. We are also grateful to Torsten Karzig, Gil Refael und Heinrich-Gregor Zirnstein for valuable discussions. This work was funded by DFG within the frame of SFB\,762 "Funktionalit\"at oxidischer Grenz\-fl\"achen". S.R. and T.M. also acknowledge the Leipzig graduate school for Natural Sciences BuildMoNa.

\bibliography{exceptional_points_in_anisotropic_microcavities}{}

\newpage

\appendix

\textbf{SUPPLEMENT}

\section{Mode computation}

\noindent\textit{Transfer matrix description}

\noindent We define the sample and polarization coordinate system as sketched in Fig.\,\ref{fig:skizze} and consider photonic states defined by the propagation angle $\theta_a$ in the ambient with respective wavevector $\vec{k}$. We investigate dispersion relations depending on the in-plane wavevector $k_{||,x}=\frac{E}{\hbar c_0}n_a\sin(\theta_a)$, where $E$ is the (real part of the) photon energy and $n_a$ is the refractive index in the ambient. The out-of-plane component of $\vec{k}$ is quantized for $m\lambda/2$-cavity photons according to $k_z=k_\perp=m\pi/d_c^\text{eff}$ with an effective cavity thickness $d_c^\text{eff}$. \\
The transverse electric polarization ($TE$) is defined by the electric field amplitude $\mathcal{E}_{TE}\,||\,y$, and the transverse magnetic one ($TM$) by the electric field amplitude $\mathcal{E}_{TE}\perp\mathcal{E}_{TM}\perp\vec{k}$ in the (isotropic) ambient. Hence, $\mathcal{E}_{TM}$ is found within the $xz$-plane. $TE$ is also referred to as $s$ ("senkrecht") polarization while $TM$ is also known as $p$ ("parallel").

We use a complex $4\times 4$ transfer matrix $\hat{T}$ according to Berreman \cite{berreman,schubertprb} which transfers incident ($\mathcal{E}^{in}$) and outgoing ($\mathcal{E}^{out}$) electric fields of the light in the basis of the directional polarization states $\mathcal{E}_{TM}$ and $\mathcal{E}_{TE}$ at both sides (i.e. $z_0$ and $z_0+d$) of a sample structure of thickness $d$:
\begin{equation}
\left ( \begin{matrix} \mathcal{E}_{TE}^{in}(z_0) \\ \mathcal{E}_{TE}^{out}(z_0) \\ \mathcal{E}_{TM}^{in}(z_0) \\ \mathcal{E}_{TM}^{out}(z_0) \end{matrix} \right ) =
\left ( \begin{matrix} T_{11} & T_{12} & T_{13} & T_{14} \\ T_{21} & T_{22} & T_{23} & T_{24} \\ T_{31} & T_{32} & T_{33} & T_{34} \\ T_{41} & T_{42} & T_{43} & T_{44} \end{matrix} \right )
\left ( \begin{matrix} \mathcal{E}_{TE}^{out}(z_0+d) \\ \mathcal{E}_{TE}^{in}(z_0+d) \\ \mathcal{E}_{TM}^{out}(z_0+d) \\ \mathcal{E}_{TM}^{in}(z_0+d) \end{matrix} \right )
\label{eq:Tintro}
\end{equation}
$\hat{T}$ contains all the structural information as layer thicknesses $d_j$ and each layer's dielectric function tensor $\hat{\varepsilon}^j$. It hence depends on the photon energy $\tilde{E}$ and the in-plane wavevector $k_{||,x}$ ($k_{||,y}=0$ assumed). 

In order to find the modes we have to apply the mode condition $\mathcal{E}^{in}=0$ while $\mathcal{E}^{out}\neq0$ \cite{felbacq,bykov,savona95unified}:
\begin{equation}
\left ( \begin{matrix} 0 \\ \mathcal{E}_{TE}^{out}(z_0) \\ 0 \\ \mathcal{E}_{TM}^{out}(z_0) \end{matrix} \right ) =
\hat{T}(\tilde{E},k_{||,x}) \cdot \left ( \begin{matrix} \mathcal{E}_{TE}^{out}(z_0+d) \\ 0 \\ \mathcal{E}_{TM}^{out}(z_0+d) \\ 0 \end{matrix} \right ) 
\label{eq:modecondition}
\end{equation}
Hence, modes, i.e. $(\mathcal{E}_{TE}^{out}, \mathcal{E}_{TM}^{out})^\text{T}$, are given by the related null eigenvectors (kernel) of a sub-matrix which represents an inverse and flipped transmission Jones matrix: 
\begin{align}
\left ( \begin{matrix} \mathcal{E}_{TE}^{out}(z_0+d) \\ \mathcal{E}_{TM}^{out}(z_0+d) \end{matrix} \right ) = \ker(\begin{pmatrix} T_{11} & T_{13} \\ T_{31} & T_{33} \end{pmatrix}) \label{eq:forward} \\
\left ( \begin{matrix} \mathcal{E}_{TE}^{out}(z_0) \\ \mathcal{E}_{TM}^{out}(z_0) \end{matrix} \right ) = \begin{pmatrix} T_{21} & T_{23} \\ T_{41} & T_{43} \end{pmatrix} \left ( \begin{matrix} \mathcal{E}_{TE}^{out}(z_0+d) \\ \mathcal{E}_{TM}^{out}(z_0+d) \end{matrix} \right )  \label{eq:backward} 
\end{align}
The modes $(\mathcal{E}_{TE}(z_0+d),\mathcal{E}_{TM}(z_0+d))^\text{T}$ and $(\mathcal{E}_{TE}(z_0),\mathcal{E}_{TM}(z_0))^\text{T}$ differ in their propagation direction with respect to $z$, i.e. they propagate towards $+z$ ("forward") or $-z$ ("backward"), respectively.  
As for a given $\vec{k}_{||}$ forward and backward traveling modes have the same energy, inverting either $k_{||,x}$ or $k_{||,y}$ is equivalent to inversion of the propagation direction along $z$ in a symmetric structure. However, this symmetry is not generally present for asymmetric systems, e.g. by considering asymmetric surroundings of the cavity layer.

\psset{unit=0.75cm}
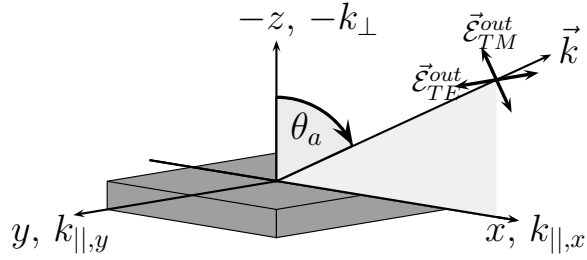
\begin{figure}
	\centering
	\begin{pspicture}[showgrid=false](-4.5,0)(5.5,5.5)
	\pspolygon*[linewidth=0.02,linecolor=mypink3!75](0,1)(0,0.5)(3,1)(3,1.5)(3,1.5)(0,2)(-3,1.5)(-3,1.0)(0,0.5)
	\pspolygon[linewidth=0.02](0,1)(0,0.5)(3,1)(3,1.5)(3,1.5)(0,2)(-3,1.5)(-3,1.0)(0,0.5)
	\psline[linewidth=0.02](0,1)(3,1.5)
	\psline[linewidth=0.02](0,1)(-3,1.5)
	\psline{->}(0,1.5)(0,4)
	\psline{<-}(-3.6,0.9)(3,2)
	\psline{->}(-2.25,1.875)(4.3,0.8)
	\put(-4.7,0.45){\large $y$, $k_{||,y}$}
	\put(3.7,0.45){\large $x$, $k_{||,x}$}
	\put(-0.7,4.2){\large $-z$, $-k_{\perp}$}
	\pspolygon*[linecolor=mypink1!40](0,1.5)(3.9,0.85)(3.9,3.3)
	\psline{->}(0,1.5)(4.875,3.75)
	\put(5,3.6){{\large $\vec{k}$}}
	\psline{->}(-2.25,1.875)(4.3,0.8)
	\psline[linewidth=0.06]{<->}(3.15,3.175)(4.65,3.425)
	\psline[linewidth=0.06]{<->}(4.17,2.715)(3.63,3.885)
	\put(3.3,4){$\vec{\mathcal{E}}_{TM}^{out}$}
	\put(2.4,3.1){$\vec{\mathcal{E}}_{TE}^{out}$}
	\pswedge[fillstyle=solid,fillcolor=mypink2!30](0,1.5){1.5}{25}{90}
	\put(0,1.5){\psarc[linewidth=1pt,arrowsize=7pt]{<-}{1.5}{25}{90}}
	\put(0.25,2.25){\large $\theta_a$}
	\end{pspicture}
	\caption{Definition of the sample coordinate system.}
	\label{fig:skizze}
\end{figure}

For a given real-valued $k_{||,x}$, the mode condition is fulfilled for complex mode energies $\tilde{E}_\text{mode}=E-i\gamma$, where $\gamma$ represents half the mode broadening ($HWHM$). \textit{Ana\-logously, it would be possible to take a real-valued energy and obtain a complex wavevector where the imaginary part is related to the mode broadening in $\vec{k}_{||}$-space.} 

In order to find the modes numerically, it is sufficient to minimize the determinant or the smallest eigenvalue of $\left ( \begin{matrix} T_{11} & T_{13} \\ T_{31} & T_{33} \end{matrix} \right )$ with respect to the complex energy. For the numerical minimization, we use a Nelder-Mead algorithm \cite{neldermead}. The respective mode (polarization) is then obtained from the related null eigenvector. 
Numerically, the matrix is only very ill-defined but not exactly singular. Hence, we have to choose the eigenvector to the vanishing eigenvalue.

\vspace{.2cm}
\noindent\textit{Mode polarization}

\noindent The modes have the structure of (flipped) Jones vectors. Applying a basis transformation they can be expressed in terms of the pseudospin or Bloch vector $\vec{\mathcal{P}}$ ($|\vec{\mathcal{P}}|=1/2$) on the surface of the Bloch sphere (SU(2)):
\begin{equation}
\vec{\mathcal{P}} = \frac{1}{2} \left (  \begin{array}{c} \mathcal{E}_{R} \\ \mathcal{E}_{L} \end{array} \right ) = 
\frac{1}{2\sqrt{2}} \begin{pmatrix} -i & 1 \\ i & 1 \end{pmatrix}
\ker( \left ( \begin{matrix} T_{11} & T_{13} \\ T_{31} & T_{33} \end{matrix} \right ) )
\label{eq:ps}
\end{equation}
Here, $\hat{T}=\hat{T}(\tilde{E}_\text{mode},k_{||,x\,\text{mode}})$ and normalized vectors of the kernel ($|\mathcal{E}_{R}|^2+|\mathcal{E}_{L}|^2=1$) are assumed. 
Except for the trivial degeneracy at $|\vec{k}_{||}|=0$ for isotropic or $c$-plane oriented uniaxial cavities where the dimension of the kernel is 2, its nullity should always be 1. If $\tilde{E}\neq\tilde{E}_\text{mode}$ it will be zero, because the matrix is not singular then. \\
$\vec{\mathcal{P}}$ corresponds directly to the normalized polarization Stokes vector $\vec{S}$ on the Poincar\'e sphere \cite{kavokinPRB92}. 
Expressing $\vec{\mathcal{P}}$ in terms of a real-valued three-dimensional vector $\vec{P}$ (SO(3) mapping), it holds:
\begin{equation}
\begin{pmatrix} P_1 \\ P_2 \\ P_3 \end{pmatrix} = 
\frac{1}{2S_0} \begin{pmatrix} S_1 \\ S_2 \\ S_3 \end{pmatrix} = 
\frac{1}{2I} \begin{pmatrix} |\mathcal{E}_{TM}|^2-|\mathcal{E}_{TE}|^2 \\ \mathcal{E}_{TM}\mathcal{E}_{TE}^*+\mathcal{E}_{TM}^*\mathcal{E}_{TE} \\ i(\mathcal{E}_{TM}\mathcal{E}_{TE}^*-\mathcal{E}_{TM}^*\mathcal{E}_{TE}) \end{pmatrix}
\label{eq:pscomponents}
\end{equation}
where $S_i$ are the components of the Stokes vector $\vec{S}$. For the intensity $I$ of the electromagnetic wave it holds $S_0=I=|\mathcal{E}_{TM}|^2+|\mathcal{E}_{TE}|^2=1$. $P_1=+\frac{1}{2}$ refers to $TM$ polarization, $P_2=+\frac{1}{2}$ to $+45^\circ$ tilted linear polarization and $P_3=+\frac{1}{2}$ to right-circular polarization (usually associated to spin down states).

\vspace{.2cm}
\noindent\textit{Anisotropy treatment and in-plane momentum space}

\noindent Cavity anisotropies are introduced to the dielectric tensor of the cavity layer $\hat{\varepsilon}_c$ which enters the transfer matrix. Generally, $\hat{\varepsilon}_c$ is complex but for transparent materials, as considered here, it is real-valued. Optical uniaxiality is represented by a tensor with two different optical constants parallel and perpendicular to the material's optical axis, $\varepsilon_{||}=n_{||}^2$ and $\varepsilon_{\perp}=n_\perp^2$, respectively:
\begin{equation}
\hat{\varepsilon}_c = 
\hat{\mathcal{R}}(\phi,\vartheta) \left ( \begin{matrix} \varepsilon_{\perp} & 0 & 0 \\ 0 & \varepsilon_{\perp} & 0 \\ 0 & 0 & \varepsilon_{||} \end{matrix} \right ) \hat{\mathcal{R}}(\phi,\vartheta)^{-1}
\label{eq:epstens}
\end{equation}
\begin{equation*} 
{\footnotesize
\hat{\mathcal{R}}(\phi,\vartheta) = \begin{pmatrix} \cos(\phi) & -\sin(\phi) & 0 \\ \sin(\phi) & \cos(\phi) & 0 \\ 0 & 0 & 1 \end{pmatrix} \begin{pmatrix} 1 & 0 & 0 \\ 0 & \cos(\vartheta) & -\sin(\vartheta) \\ 0 & \sin(\vartheta) & \cos(\vartheta) \end{pmatrix}
}
\end{equation*} 
For arbitrary orientation of the optical axis, Euler rotation matrices $\hat{\mathcal{R}}(\phi,\vartheta)$ are applied \cite{goldstein,schubertprb}. The rotations are carried out as follows: 1. $\phi$ rotation about $z$ (mathematically positive in the $xy$-plane), 2. $\vartheta$ rotation about $x'$ which resulted from the first rotation (mathematically positive in the $y'z$-plane). The medium's optical axis is aligned in the $xy$ plane for $\vartheta=\pm90^\circ$. \\
For coverage of the entire in-plane momentum space, we use rotations about $\tilde{\phi}$ and $\tilde{\vartheta}$ to transfer the laboratory $xy$ coordinates to $\tilde{x}\tilde{y}$ (and still calculate modes for $k_{||,x}$). 
For a given orientation of the optical axis ($\phi,\,\vartheta$) and an arbitrary in-plane wavevector described by $\vec{k}_{||}=(k_{||,x},k_{||,y})^\text{T}=|\vec{k}_{||}|(\cos(\varphi),\sin(\varphi))^\text{T}$, 
the Euler angles necessary for construction of the transfer matrix $\hat{T}$ are then given by
\begin{equation}
\tilde{\phi}=\phi-\varphi,\qquad\tilde{\vartheta}=\vartheta
\end{equation}
It holds $\tilde{\phi}=-\varphi$ for the optical axis along the $y$-direction ($\phi=0$, $\vartheta=90^\circ$). \\
In order to describe the linear polarization components with respect to the laboratory coordinate system ($\mathcal{E}_{x}$, $\mathcal{E}_{y}$) instead of using $\mathcal{E}_{TE}$ and $\mathcal{E}_{TM}$, also the obtained pseudospin $\vec{\mathcal{P}}$ needs to be modified. We define $\vec{\tilde{\mathcal{P}}}$ as follows: 
\begin{equation}
\vec{\tilde{\mathcal{P}}} = \begin{pmatrix} 1& 0 \\ 0 & e^{\pm2i\varphi} \end{pmatrix} \vec{\mathcal{P}}
\label{eq:psneu1}
\end{equation}
In $\vec{\tilde{\mathcal{P}}}$, the $TE$ and $TM$ fields are tilted into the $xy$-plane as it would be observed experimentally in Fourier space imaging. 
Again, the polarization vector can be expressed by three real-valued components as $\vec{\tilde{P}}$:
\begin{equation}
\begin{pmatrix} \tilde{P}_1 \\ \tilde{P}_2 \\ \tilde{P}_3 \end{pmatrix} = 
\begin{pmatrix} \cos(2\varphi) & \mp\sin(2\varphi) & 0 \\ \pm\sin(2\varphi) & \cos(2\varphi) & 0 \\ 0 & 0 & 1 \end{pmatrix}
\begin{pmatrix} P_1 \\ P_2 \\ P_3 \end{pmatrix}
\label{eq:psneu2}
\end{equation}
where the upper signs of $\pm$ and $\mp$ are valid for forward traveling waves (towards $+z$) and the lower ones for backwards (towards $-z$) traveling modes (cf. equations \ref{eq:forward} and \ref{eq:backward}). 
Now, $\tilde{P_1}=+\frac{1}{2}$ refers to linear polarization along the $x$ axis of the laboratory coordinate system, $\tilde{P_1}=-\frac{1}{2}$ to linear polarization along the $y$ axis.

\section{Construction of the transfer matrix}

\noindent The transfer matrix according to Berreman \cite{berreman} for a slab of $N$ layers depends on the photon energy $E$ and in-plane wavevector $k_{||,x}$ and reads in the formulation of Schubert \cite{schubertprb} with $k_{||,x}=\frac{E}{\hbar c_0}n_a\sin\theta_a$ (ambient refractive index $n_a$, isotropic ambient and substrate, propagation angle $\theta_a$ in the ambient):
\begin{equation}
\hat{T} = \hat{T}_{I/O}^{-1}\prod_{j=1}^Ne^{-i\frac{Ed_j}{\hbar c_0}\hat{\Delta}_j}\hat{T}_{I/O}~,
\label{eq:transfermatrix}
\end{equation}
with layer thicknesses $d_j$, reduced Planck constant $\hbar$, vacuum speed of light $c_0$. 
The in-/out coupling matrices $\hat{T}_{I/O}$ transfer the for- and backward traveling electric field amplitudes in $TE$ and $TM$ (eigenmode) basis, i.e. $(\mathcal{E}_{TE}^\rightarrow, \mathcal{E}_{TE}^\leftarrow, \mathcal{E}_{TM}^\rightarrow, \mathcal{E}_{TM}^\leftarrow)^\text{T}$, in a medium $n_a$ to the tangential electric and magnetic field components $(\mathcal{E}_{x}, \mathcal{E}_{y}, \mathcal{H}_{x}, \mathcal{H}_{y})^\text{T}$ in the $xy$-plane. Here, "$\rightarrow$" means propagation towards $+z$ and "$\leftarrow$" towards $-z$. 
It is useful to introduce the auxiliary wavevector projection $\tilde{k}=:n_j\sin{\theta_j}=k_{||,x}\frac{\hbar c_0}{E}$ which is a constant along all media $j$: 
\begin{equation}
\hat{T}_{I/O} = \begin{pmatrix} 0 & 0 & \frac{\sqrt{n_a^2-\tilde{k}^2}}{n_a} & -\frac{\sqrt{n_a^2-\tilde{k}^2}}{n_a} \\ 1 & 1 & 0 & 0 \\ -\sqrt{n_a^2-\tilde{k}^2} & \sqrt{n_a^2-\tilde{k}^2} & 0 & 0 \\ 0 & 0 & n_a & n_a \end{pmatrix} 
\label{eq:tio}
\end{equation}
and 
\begin{equation}
\hat{T}_{I/O}^{-1} = \frac{1}{2} 
\left ( \begin{matrix} 0 & 1 & \frac{-1}{\sqrt{n_a^2-\tilde{k}^2}} & 0 \\ 0 & 1 & \frac{1}{\sqrt{n_a^2-\tilde{k}^2}} & 0 \\ \frac{n_a}{\sqrt{n_a^2-\tilde{k}^2}} & 0 & 0 & \frac{1}{n_a} \\ \frac{-n_a}{\sqrt{n_a^2-\tilde{k}^2}} & 0 & 0 & \frac{1}{n_a} \end{matrix} \right )\,
\label{tioinv}
\end{equation}
In Eq.\,\ref{eq:transfermatrix}, $n_a$ in $\hat{T}_{I/O}^{-1}$ represents the ambient (superstrate) medium while $n_a$ in $\hat{T}_{I/O}$ represents the substrate medium. 
Those definitions of $\hat{T}_{I/O}$ imply a right-handed polarization coordinate systems where the $x$-axis is inverted for backward compared to forward traveling waves: $x^\leftarrow_{pol}=-x^\rightarrow_{pol}$ and $y^\leftarrow_{pol}=y^\rightarrow_{pol}$. 

The transfer through each layer $j$ is represented by matrix exponentials called partial transfer matrices $e^{-ik_0d_j\hat{\Delta}_j}$. The matrix $\hat{\Delta}$ describes the differential wave propagation along $z$ for the $j$\textsuperscript{th} layer in terms of the dielectric tensor which is given by a complex symmetric $3\times 3$ matrix $\hat{\varepsilon}^j$ for the respective layer and generally depends on the photon energy $\tilde{E}$:
\begin{align}
\hat{\Delta}_j = 
{\footnotesize \left ( \begin{matrix} -\tilde{k}\frac{\varepsilon^j_{31}}{\varepsilon^j_{33}} & -\tilde{k}\frac{\varepsilon^j_{32}}{\varepsilon^j_{33}} & 0 & 1-\frac{\tilde{k}^2}{\varepsilon^j_{33}} \\ 0 & 0 & -1 & 0 \\ \frac{\varepsilon^j_{23}\varepsilon^j_{31}}{\varepsilon^j_{33}}-\varepsilon^j_{21} & \tilde{k}^2-\varepsilon^j_{22}+\frac{\varepsilon^j_{23}\varepsilon^j_{32}}{\varepsilon^j_{33}} & 0 & \tilde{k}\frac{\varepsilon^j_{23}}{\varepsilon^j_{33}} \\ \varepsilon^j_{11}-\frac{\varepsilon^j_{13}\varepsilon^j_{31}}{\varepsilon^j_{33}} & \varepsilon^j_{12}-\frac{\varepsilon^j_{13}\varepsilon^j_{32}}{\varepsilon^j_{33}} & 0 & -\tilde{k}\frac{\varepsilon^j_{13}}{\varepsilon^j_{33}} \end{matrix} \right ) }
\label{eq:delta}
\end{align}

In fact, the columns of $\hat{T}_{I/O}$ are eigenvectors of $\hat{\Delta}$ for the ambient medium with coefficients of the eigenvectors chosen in order to fulfill the geometric constraints which hold for projection of the $TE$/$TM$ fields.

\section{Mode condition without restricting the polarization beforehand}

\noindent Considering the electric fields inside the cavity layer yields the following mode condition for the $TE$/$TM$ fields at the boundaries of the cavity layer ($z=z_{c0}$ and $z=z_{c0}+d_c$, respectively): 
\begin{align}
& \begin{pmatrix} 0 & r_{ss}^c & 0 & r_{ps}^c \\ 0 & 1 & 0 & 0 \\ 0 & r_{sp}^c & 0 & r_{pp}^c \\ 0 & 0 & 0 & 1 \end{pmatrix}
\begin{pmatrix} \mathcal{E}_{TE}^{in}(z_{c0}) \\ \mathcal{E}_{TE}^{out}(z_{c0}) \\ \mathcal{E}_{TM}^{in}(z_{c0}) \\ \mathcal{E}_{TM}^{out}(z_{c0})
\end{pmatrix} \label{eq:modeconditionrigoros} \\
& =
\hat{\tilde{T}}_{I/O}^{-1} e^{-i\frac{E}{\hbar c_0}d_c\hat{\Delta}_c}\hat{\tilde{T}}_{I/O}
\begin{pmatrix} 1 & 0 & 0 & 0 \\ r_{ss}^c & 0 & r_{ps}^c & 0 \\ 0 & 0 & 1 & 0 \\ r_{sp}^c & 0 & r_{pp}^c & 0 \end{pmatrix}
\begin{pmatrix} \mathcal{E}_{TE}^{out}(z_{c0}+d_c) \\ \mathcal{E}_{TE}^{in}(z_{c0}+d_c) \\ \mathcal{E}_{TM}^{out}(z_{c0}+d_c) \\ \mathcal{E}_{TM}^{in}(z_{c0}+d_c)\end{pmatrix}  \nonumber
\end{align}
Here, $r_{ij}^c$ are the polarization-dependent complex reflection coefficients for light inside the anisotropic cavity layer which is reflected off the DBRs. $\hat{\tilde{T}}_{I/O}$ is different from Eqs.\,\ref{eq:tio} and \ref{tioinv} as ordinary and extra-ordinary refractive indices of the cavity medium need to be distinguished. 
For (pseudo-)isotropic cases, $\hat{\tilde{T}}_{I/O}$ is similar to $\hat{T}_{I/O}$ (cf. Eq.\,\ref{eq:tio}) and $\hat{T}_{I/O}^{-1} e^{-i\frac{E}{\hbar c_0}d_c\hat{\Delta}_c}\hat{T}_{I/O}$ becomes a diagonal matrix. 
Hence, no cross-polarized reflection occurs ($r_{ps}^c=r_{sp}^c=0$) and the problem can be treated independently for $TE$ and $TM$ polarization and the well known isotropic mode condition $r_{p/s}^{c}e^{i\frac{E}{\hbar c_0}d_cn_c^{p/s}\cos{\theta_c}}=\pm1$ can be deduced from Eq.\,\ref{eq:modeconditionrigoros} likewise for both polarizations \cite{mcbook,sturm2011}. 
For arbitrarily oriented anisotropic cavity layers, both, determination of $r_{ij}^c$ and $\hat{\tilde{T}}_{I/O}$ become very complicated, because the eigenmodes do not represent $TE$ and $TM$ polarizations. 
Hence, it is more convenient to consider the fields outside the resonator structure. For the electric fields at the top $z=z_0$ which are transmitted through the top DBR, we get:
\begin{align}
\begin{pmatrix} \mathcal{E}_{TE}^{in}(z_0) \\ \mathcal{E}_{TE}^{out}(z_0) \\ \mathcal{E}_{TM}^{in}(z_0) \\ \mathcal{E}_{TM}^{out}(z_0) \end{pmatrix} 
= 
\begin{pmatrix} 0 & 0 & 0 & 0 \\ 0 & t^c_{ss} & 0 & t^c_{ps} \\ 0 & 0 & 0 & 0 \\ 0 & t^c_{sp} & 0 & t^c_{pp} \end{pmatrix}
\begin{pmatrix} \mathcal{E}_{TE}^{in}(z_{c0}) \\ \mathcal{E}_{TE}^{out}(z_{c0}) \\ \mathcal{E}_{TM}^{in}(z_{c0}) \\ \mathcal{E}_{TM}^{out}(z_{c0})
\end{pmatrix}
= 
\begin{pmatrix} 0 \\ \mathcal{E}_{TE}^{out}(z_0) \\ 0 \\ \mathcal{E}_{TM}^{out}(z_0) \end{pmatrix}
\label{eq:dbrtransfer}
\end{align}
where the complex transmission coefficients $t^c_{ij}$ hold for light traveling through the DBR towards $z_0$ (backward propagating modes). A transfer similar to Eq.\,\ref{eq:dbrtransfer} is valid for forward propagating waves and the fields at the bottom side $z=z_0+d$. 
Hence we end up with the mode condition introduced by Eq.\,\ref{eq:modecondition}: $\mathcal{E}_{TE/TM}^{in}= 0$ with $\mathcal{E}_{TE/TM}^{out}\neq 0$. This is the generalization of the condition introduced by Savona \textit{et al.} \cite{savona95unified}.

In an even more general approach it is straight forward to rewrite the transfer matrix $\hat{T}$ and Eq.\,\ref{eq:Tintro} in terms of a scattering matrix $\hat{S}$:
\begin{equation}
\left ( \begin{matrix} \mathcal{E}_{TM}^{out}(z_0) \\ \mathcal{E}_{TE}^{out}(z_0) \\ \mathcal{E}_{TM}^{out}(z_0+d) \\ \mathcal{E}_{TE}^{out}(z_0+d) \end{matrix} \right ) =
\hat{S} \cdot \left ( \begin{matrix} \mathcal{E}_{TM}^{in}(z_0) \\ \mathcal{E}_{TE}^{in}(z_0) \\ \mathcal{E}_{TM}^{in}(z_0+d) \\ \mathcal{E}_{TE}^{in}(z_0+d) \end{matrix} \right ) ~,
\end{equation}
where the block matrices of $\hat{S}$ can be considered as the respective unnormalized Jones matrices $\hat{J}$ consisting of the complex reflection and transmission coefficients $r_{ij}$ and $t_{ij}$ for the structure. Here, $\hat{J}$ represent the Jones matrices for the front side and $\hat{\tilde{J}}$ for the backside with respect to incoming light:
\begin{equation}
\hat{S} = \left ( \begin{array}{cc|cc} r^{pp} & r^{sp} & \tilde{t}^{pp} & \tilde{t}^{sp} \\ r^{ps} & r^{ss} & \tilde{t}^{ps} & \tilde{t}^{ss} \\ \hline t^{pp} & t^{sp} & \tilde{r}^{pp} & \tilde{r}^{sp} \\ t^{ps} & t^{ss} & \tilde{r}^{ps} & \tilde{r}^{ss} \end{array} \right ) 
= \left ( \begin{array}{c|c} \hat{J}_{refl} & \hat{\tilde{J}}_{trans} \\ \hline \hat{J}_{trans} & \hat{\tilde{J}}_{refl} \end{array} \right ) 
\end{equation}

Now, the mode condition $\mathcal{E}_{TE/TM}^{in}= 0$ along with $\mathcal{E}_{TE/TM}^{out}\neq 0$ is fulfilled if $\hat{S}^{-1}$ becomes singular and modes are given as the nullspace $\ker(\hat{S}^{-1})$. 
$\hat{S}$ and $\hat{S}^{-1}$ are holomorphic and in the limit $\tilde{E}\rightarrow\tilde{E}_\text{mode}$ it holds $\det(\hat{S})=1/\det(\hat{S}^{-1})$. Hence, the singularities of $\hat{S}^{-1}$ are poles of $\hat{S}$ where $|\det(\hat{S})|$ approaches infinity \cite{felbacq,bykov}. 
This is also true for each individual Jones matrix, i.e. block matrix of $\hat{S}$.

\section{Further mode polarization patterns}

\begin{figure*}
\centering
\includegraphics[width=\textwidth]{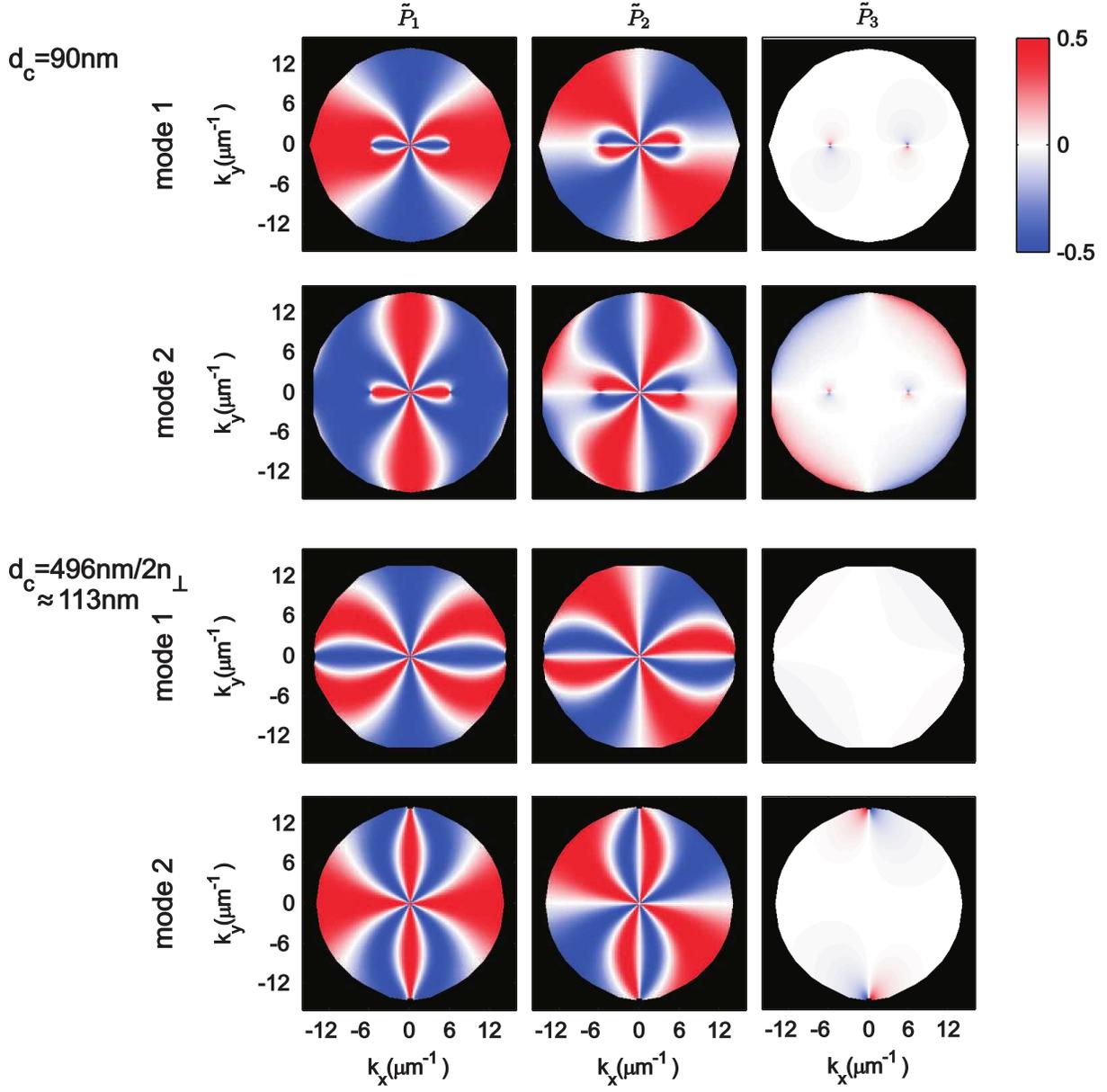}
\caption{Mode polarization in terms of the pseudospin components for a microcavity with optically uniaxial cavity layer, the optical axis of which is aligned along $y$ (cf. Fig.\,2 in the main text). Two different cavity thicknesses $d_c$ are shown. For comparison, Fig.\,2 in the main text depicts $d_c=130$\,nm.}
\label{fig:polarisationsmuster}
\end{figure*}

\noindent As discussed in the main text, the occurrence of the exceptional points can be controlled by varying the cavity thickness $d_{c}$. Figure \ref{fig:polarisationsmuster} shows the polarization in addition to Fig.\,2 in the main text for two further values of $d_c$. 
The features of the polarization patterns are the same if the optical cavity thickness $n_cd_c$ ($n_c=n_{\perp,||}$) is smaller or larger than half the DBR central wavelength $\lambda_\text{DBR}$. Only, the degeneracies are shifted to other $\vec{k}_{||}$ values. 
If $n_cd_c\approx\lambda_\text{DBR}/2$, the degeneracies are pushed out of the light cone. However, circular polarization contributions can still be identified at the edge of the light cone. Remarkably, if $n_cd_c\not\approx\lambda_\text{DBR}/2$ such circular polarization traces are only pronounced for the narrower mode, i.e. the one with the smaller $\gamma$.

\end{document}